\documentclass[11pt]{article}
\usepackage{amsmath}
\usepackage{amssymb}
\usepackage{graphicx}
\numberwithin{equation}{section} \textwidth=160mm
\begin{document}
\setcounter{page}{0} \thispagestyle{empty}
\begin{flushright}
\end{flushright}
\vspace*{2.0cm}
%
%
%
%
\begin{center}
{\large\bf Electrodynamics in an LTB scenario}
\end{center}
\vspace*{1cm}
\renewcommand{\thefootnote}{\fnsymbol{footnote}}
\begin{center}
G. Fanizza$^{1,2}$\protect\footnote{Electronic address:
              {\tt giuseppe.fanizza@ba.infn.it}},
and
L. Tedesco$^{1,2}$\protect\footnote{Electronic address:
              {\tt luigi.tedesco@ba.infn.it}}\\[0.5cm]

$^1${\em Dipartimento di Fisica, Universit\`a di Bari, I-70126 Bari, Italy}\\[0.2cm]
$^2${\em INFN - Sezione di Bari, I-70126 Bari, Italy}
\end{center}

\date{\today}

\begin{abstract}
In this article we analyze the electrodynamics in curved space-time in the Lema\^itre-Tolman-Bondi metric. We calculate the most general scale factor in this inhomogeneous Universe. We also study the presence of electromagnetic field bubbles in the Universe.
\end{abstract}
\newpage
\section{Introduction}
The Universe is described in  models, the so called Friedmann-Lema\^itre-Robertson-Walker (FLRW) models, in which it is homogeneous and isotropic and the matter is a gas of particles with a pressure $p>$ 0. According to the cosmological principle our Universe is homogeneous and isotropic when we consider scales larger that a few hundred Mpc. A very intriguing question here is if there is a scale above which the Universe is approximately FLRW. These models were developed in the period  1917-1935 by de Sitter, Friedmann, Lema\^itre, Robertson, Walker and Einstein.
The main reason to consider an FLRW model is that it is universally recognized as a very good approximation to a more realistic description of our cosmos, but it is essential to remember that the cosmological principle is postulate not a law of nature. 
\\
During the last decade a great effort has been made in understanding that the Universe is locally far from homogeneity (see for example the formation of non-linear structure). The first study of the effect of inhomogeneity and anisotropy in the Universe is known as `the fitting problem' \cite{ellis} and the inhomogeneities are able to  explain the accelerating expansion of the Universe \cite{{rasanen}, {chuang}, {paranjape}, {kozaki}, {rasanen2},{enqvist},{cosmai}}. 
The inhomogeneous models of the Universe are to be able to explain observational results due to the so called `dark energy' and they are just these effects of inhomogeneities that may  mimic the accelerated expansion of the Universe \cite{cosmai} and then remove the need to postulate the dark energy.
\\
The inhomogeneity of the Universe is a very intriguing open research field and intensive study has been obtained \cite{{pietronero}, {joyce}, {sylos},{labini}}. Among the structures in the Universe we have found very large voids that dominate the Universe, while matter is mostly  distributed in filamentary structures that surround the voids. When we consider larger scales we see very strange structures such as the so called `Great Wall'. Therefore it is important to consider photons that trade in these voids and to study the differences with photons that travel in a homogeneous and isotropic Universe as an FLRW model. These studies started some years ago \cite {{zeld, bertotti, dash, gunn, kantowski, dyer, weinberg}}.
\\
\\
In the last years many papers have been published on the problem of electrodynamics in an isotropic and homogeneous gravitational background in FLRW models. 
The electrodynamics effect in an anisotropic Universe has been studied in \cite{Ciarcelluti:2012pc}, the author studies very interesting astrophysical consequences connected with anisotropic expansion of the Universe. In particular he suggests the appearance of polarization of electromagnetic radiation when it passes through local anisotropic regions.
\\
The purpose of this article is to analyze the electrodynamics in LTB models, that is to say, we study electrodynamics in curved space-time and in particular we find the most general scale factor in this background.
\\
The paper is structured as follows. In Sect.~2 we tidy the electromagnetic field in curved space-time in a LTB background metric. We obtain the components of the electromagnetic energy momentum and we calculate the `new' metric taking into account all contributions. In Sect.~3 we study the Maxwell equations in curved space-time and we solve the Einstein equations. It is possible to obtain the most general energy density of the electromagnetic field. In Sect.~4 we solve the radial-radial component of the Einstein equation in order to have the most general scale factor in this context. In Sect. 5 we consider the case of electromagnetic field bubbles in the Universe. The discussion of the results and final remarks are presented  in Sect.~6.     
\\
\\
\\
\section{Lema\^itre-Tolman-Bondi metric with electromagnetic field}
In this section we study electrodynamics and we calculate the energy-momentum tensor for the electromagnetic field in a inhomogeneous metric. In fact
in order to describe the inhomogeneities, let us consider the well-studied Lema\^itre-Tolman-Bondi metric:
\begin{equation}
\label{eq:LTB}
ds^2=dt^2-X^2(t,r)\,dr^2-A^2(t,r)\left[ d\theta^2+\sin^2\theta\,d\phi^2 \right]
\end{equation}
with $\sqrt{-g}=X(t,r)A^2(t,r)\sin\theta$. At the present step, there is no relation between $X(t,r)$ and $A(t,r)$.  Let us introduce electrodynamics in curved space-time. To this end, a very interesting pedagogical introduction to formulate electrodynamics in curved space-time may be founded in \cite{Subramanian:2009fu} in which the author also analyzed the Maxwell equations for an expanding Universe with the metric of a spatially flat Friedmann-Robertson-Walker space-time.    The electromagnetic field tensor is $F_{\mu \nu} = A_{\nu ; \mu} - A_{\mu ; \nu} = A_{\nu, \mu} - A_{\mu, \nu}$ where the four-potential is $A_{\mu}$. Following \cite{Subramanian:2009fu} it is possible to obtain for the electromagnetic field tensor and its dual the following expressions:
\begin{align}
&F_{\mu\nu}=u_\mu E_\nu-u_\nu E_\mu+\epsilon_{\mu\nu\rho\sigma}B^\rho u^\sigma\\
&^*F^{\mu\nu}=\frac{1}{2}\,\epsilon^{\mu\nu\alpha\beta}F_{\alpha\beta} = \epsilon^{\mu \nu \alpha \beta} u_{\alpha} E_{\beta} + u^{\mu} B^{\nu} - B^{\mu} u^{\nu}  
\end{align}
where $u_\mu$ is the four-velocity of the observer, $E_\nu$ and $B^\rho$ are, respectively, the electric and the magnetic field, $\epsilon_{\mu\nu\rho\sigma}=\sqrt{-g}\,\mathcal{A}_{\mu\nu\rho\sigma}$ is the totally covariant antisymmetric tensor, $\epsilon^{\mu\nu\rho\sigma}=\mathcal{A}^{\mu\nu\rho\sigma}/\sqrt{-g}$ is the controvariant one, and $\mathcal{A}_{\mu\nu\rho\sigma}$ is the Levi-Civita symbol with $\mathcal{A}_{0123}=1$ (+ (-) for any even (odd) permutations of 0123). 
\\
The Maxwell equations in a curved space-time:
\begin{align}
\label{eq:MaxwellEquation}
\nabla_\nu& F^{\mu\nu}=4\pi\,J^\mu\\
\label{eq:dualMaxwellEquation}
\nabla_\nu& \,^*F^{\mu\nu}=0
\end{align}
where $\nabla_\nu$ is the covariant derivative and $J^{\mu}=(\rho, \vec{J})$ is the electromagnetic four-current density. The components of the electromagnetic field in the LTB metric are
\begin{equation}
\label{compnent}
F^{0l} = E^{(l)} \,\;\;\;\; F^{12} = \frac{\sin \theta} {X} B^{(3)} \;\;\;\;\; F^{13} = - \frac {B^{(2)}} {X sin \theta} \;\;\;\;\; F^{23} = \frac{X} {A^2 \sin \theta} B^{(1)} \, ,
\end{equation}
\begin{equation}
\label{dualcompnent}
^*F^{0l} = B^{(l)} \,\;\;\;\; ^*F^{12} = - \frac{\sin \theta} {X} E^{(3)} \;\;\;\;\; ^*F^{13} =  \frac {E^{(2)}} {X \sin \theta} \;\;\;\;\; ^*F^{23} = \frac{X} {A^2 \sin \theta} E^{(1)}.
\end{equation}
The presence of an electromagnetic field contributes to the energy-momentum tensor with the following term:
\begin{equation}
\label{eqTem}
T_{\mu\nu}^\text{(EM)}=-F_{\mu\alpha}F_{\nu\beta}\,g^{\alpha\beta}+\frac{1}{4}\left( F_{\alpha\beta}F^{\alpha\beta}+J^\alpha\tilde A_\alpha\right)\,g_{\mu\nu}
\end{equation}
For our purposes, we neglect the coupling term $J^\alpha\tilde A_\alpha$; in fact, according to the ordinary cosmology, the greatest part of the matter is dark i.e. electrically neutral. In this way, the eventually appearing fluctuations of charge in the barotropic fluid are due to the ordinary matter and so it contributes with a second order correction that we neglect.
At this step, we are able to write the non-null component of $T_{\mu\nu}^\text{(EM)}$ for a comoving observer ($u_\mu=(1,\vec0)$) as follows:
\begin{align}
\label{Tensor}
&T_{00}^\text{(EM)}=X^2(E^{(1)})^2+A^2(E^{(2)})^2+A^2\sin^2\theta(E^{(3)})^2+\frac{1}{4}\,F_{\alpha\beta}F^{\alpha\beta}\\
&T_{11}^\text{(EM)}=-X^4(E^{(1)})^2+X^2A^2\sin^2\theta(B^{(3)})^2+X^2A^2(B^{(2)})^2-\frac{X^2}{4}\, F_{\alpha\beta}F^{\alpha\beta}\\
&T_{22}^\text{(EM)}=T_{33}^\text{(EM)}=-A^4(E^{(2)})+A^4\sin^2\theta(B^{(3)})^2+X^2A^2(B^{(1)})^2-\frac{A^2}{4}\, F_{\alpha\beta}F^{\alpha\beta}\\
&T_{10}^\text{(EM)}= X A^2 \sin \theta \,(E^{(3)} B^{(2)} - E^{(2)} B^{(3)})\\
&T_{20}^\text{(EM)}= X A^2 \sin \theta \,(E^{(1)} B^{(3)} - E^{(3)} B^{(1)})\\
&T_{30}^\text{(EM)}= X A^2 \sin \theta \,(E^{(2)} B^{(1)} - E^{(1)} B^{(2)})\\
&T_{12}^\text{(EM)}= - X^2 A^2 (E^{(1)} E^{(2)} + B^{(1)} B^{(2)})\\
&T_{13}^\text{(EM)}= - X^2 A^2 \sin^2 \theta \, (E^{(1)} E^{(3)} + B^{(1)} B^{(3)})\\
&T_{23}^\text{(EM)}= -  A^4 \sin^2 \theta \, (E^{(2)} E^{(3)} - B^{(2)} B^{(3)}). \\
%
\end{align}
We find that the energy-momentum tensor is not diagonal. Let us consider the Einstein equation $G_{\mu \nu} = 8 \pi G_N T{\mu \nu}$, where $G_{\mu \nu}$ is the Einstein tensor, $G_N$ is the Newton constant. The only non-null off-diagonal term in $G_{\mu \nu}$ for the metric Eq.~(\ref{eq:LTB}) is
\begin{equation}
G_{10}=2\left( \frac{A'}{A}\frac{\dot X}{X}-\frac{\dot A'}{A'} \right)
\end{equation}
with $'\equiv\partial/\partial r$ and $\dot{}\equiv\partial/\partial t$. Therefore the Einstein equations $G_{\mu \nu} = 8\,  \pi \, G \, T_{\mu \nu}$ impose us the following equalities:
\begin{equation}
T_{20}^\text{(EM)}=T_{30}^\text{(EM)}=T_{12}^\text{(EM)}=T_{13}^\text{(EM)}=T_{23}^\text{(EM)}=0\, ,
\end{equation}
so that, taking into account  Eqs. (2.13)-(2.18), a possible solution is given by $B^2=B^3=0$ and $E^2=E^3=0$. In this way we also have  $T_{10}^\text{(EM)}=0\Rightarrow G_{10}=0$ i.e.
\begin{equation}
X(t,r)=\frac{A'(t,r)}{\sqrt{1-k( r)}}
\end{equation}
where $k( r)$ is the spatial curvature as is well-known in the matter dominated case. Our new metric becomes 
\begin{equation}
\label{neweq}
ds^2 = dt^2 - \frac {{A'}^2 (t,r)} {1-k(r)} \, dr^2 - A^2(t,r) \, d \Omega^2
\end{equation}
as in the case without electromagnetic field.
\\
\\
\\
\\
\section{Electrodynamic's equations}
In this section we want to study the Maxwell equations in a curved space-time:
\begin{align}
\label{eq:MaxwellEquation}
\nabla_\nu& F^{\mu\nu}=4\pi\,J^\mu\\
\label{eq:dualMaxwellEquation}
\nabla_\nu& \,^*F^{\mu\nu}=0
\end{align}
In order to calculate the Maxwell equations it is important to stress that our fields are 
\begin{equation}
\label{eqefield}
E^{\mu} = (0, E^{(1)} (t, r, \theta, \phi), 0, 0)
\end{equation}
\begin{equation}
\label{eqbfield}
B^{\mu} = (0, B^{(1)} (t, r, \theta, \phi), 0, 0).
\end{equation}
Now it is useful to make explicit the cases $\mu=0,1,2,3$ for the Maxwell equations (\ref{eq:MaxwellEquation}) and (\ref{eq:dualMaxwellEquation}). 
\begin{align}
\label{eq:Er}
\partial_1(XA^2E^{(1)})&=4\pi XA^2J^0\\
\label{eq:Et}
\partial_0(XA^2E^{(1)})&=-4\pi XA^2J^1\\
\label{eq:Bfi}
X\partial_3B^{(1)}&=4\pi A^2\sin\theta J^2\\
\label{eq:Btheta}
X\partial_2B^{(1)}&=-4\pi A^2\sin^2\theta J^3\\
\label{eq:Br}
\partial_1(XA^2B^{(1)})&=0\\
\label{eq:Bt}
\partial_0(XA^2B^{(1)})&=0\\
\label{eq:Efi}
\partial_3E^{(1)}&=0\\
\label{eq:Etheta}
\partial_2E^{(1)}&=0.
\end{align}
Our goal is to solve the two Einstein equations, 
\begin{equation}
\label{einst00}
G_0^{\,\,0} = 8 \pi G ({T_0^{\,\, 0}}^{\text{(M)}}+ {T_0^{\,\, 0}}^{\text{(EM)}})
\end{equation}
and
\begin{equation}
\label{einst11}
G_1^{\,\,1} = 8 \pi G ({T_1^{\,\, 1}}^{\text{(M)}}+ {T_1^{\,\, 1}}^{\text{(EM)}})
\end{equation}
where 
\begin{equation}
\label{Tmateria}
{T_{\mu}^{\,\,\nu}}^{\text{(M)}} = \rho_m u_{\mu} u^{\nu} = \rho_m \delta_{\mu 0} \delta^{\nu 0}
\end{equation}
where $\rho_m$ is the energy density of the pressureless fluid of matter and ${T_{\mu}^{\,\, \nu}}^{\text{(EM)}}$ is given by Eq.~(\ref{eqTem}). We must take into account that the only two independent Einstein equations are
\begin{align}
\label{eqG00}
\frac{\dot A^2+k}{A^2}+\frac{2\dot A\dot A'+k'}{AA'}&=8\pi G\left[ \rho_m+\frac{A'^2(E^2+B^2)}{2\,(1-k)} - J^{\alpha} A_{\alpha} \right]\\
\label{eqG11}
\frac{\dot A^2+2A\ddot A+k}{A^2}&=8\pi G\left[\frac{A'^2(E^2+B^2)}{2\,(1-k)} - J^{\alpha} A_{\alpha} \right]
\end{align}
that respectively correspond to the time-time equation and the radial-radial one Eqs. (\ref{eqG00}) and (\ref{eqG11}). From now on we assume that matter and field do not interact in order to eliminate the term $J^{\alpha} A_{\alpha}$ and  in order to simplify the notation, we have denoted electrical and magnetic fields without superscript. In order to solve Eqs.~(\ref{eqG00}) and (\ref{eqG11}) it is necessary to have the expression for the fields $E$ and $B$. 
\\
From \eqref{eq:Efi} and \eqref{eq:Etheta} we see that $E$ is independent from $\theta$ and $\phi$. By integration of \eqref{eq:Er} we obtain
\begin{equation}
\label{eq:Efield1}
E(t,r)=\frac{\epsilon(t)+\chi(t,r)}{X(t,r)A^2(t,rt)}
\end{equation}
with
\begin{equation}
\chi(t,r)\equiv4\pi\int_0^rX(t,\bar r)A^2(t,\bar r)J^0(t,\bar r)d\bar r.
\end{equation}
This expression is related to the radial charged current by the Eq.~\eqref{eq:Et}, in fact substituting Eq.~(\ref{eq:Efield1}) in eq. (3.6) we have
\begin{equation}
\dot \epsilon+\dot \chi=-4\pi XA^2J^{(1)}.
\end{equation}
However, the charged matter belongs to the comoving matter; this means that the four-current appears as $J^\alpha=(J^0,\vec 0)$. In such a way
\begin{equation}
\label{eq:Efield2}
\dot \epsilon+\dot \chi=0\Rightarrow \epsilon(t,r)+\chi(t,r)\equiv \epsilon_0+\chi_0( r).
\end{equation}
At the same way, because the current is null, also we see that the magnetic field depends only by $t$ and $r$ (Eq. \eqref{eq:Bfi}, Eq. \eqref{eq:Btheta}) and these dependences are fixed by Eqs.~\eqref{eq:Br} and \eqref{eq:Bt} as follows:
\begin{equation}
\label{eq:Bfield}
B(t,r)=\frac{\beta_0}{X(t,r)A(t,r)^2}.
\end{equation}
Equations \eqref{eq:Efield1}, \eqref{eq:Efield2} and \eqref{eq:Bfield} allow us to write the energy density of the electromagnetic field as follows:
\begin{equation}
\label{eq:totalDensity}
E^2+B^2=\frac{\left[ \epsilon_0+\chi_0( r) \right]^2+\beta_0^2}{X^2A^4}\equiv\frac{\gamma(r)}{4\pi GX^2A^4}.
\end{equation}
The last expression in the third member of Eq.~(\ref{eq:totalDensity}) is written in order to simplify the equations in the next section.
\\
\\
\\
\section{Scale factor}
In this section we study the solution of the Einstein equations in the presence of electromagnetic field given by Eq.~(\ref{eq:totalDensity}).
Now let us rewrite Eq. (\ref{eqG11}) with the aim of the Eq.~\eqref{eq:totalDensity}
\begin{equation}
\dot A^2+2A\ddot A+k( r)=\frac{\gamma( r)}{A^2}
\end{equation}
and multiply by $\dot A$
\begin{equation}
\dot A^3+2A\dot A\ddot A=-k(r)\dot A+\gamma( r)\frac{\dot A}{A^2}.
\end{equation}
The left side of the last equation can be expressed as $\partial_0(A\dot A^2)$; a first integration allows us to obtain the following equalities:
\begin{multline}
\label{eq:fieldEquation}
A\dot A^2=-k(r) A+\alpha(r)+\int\frac{\gamma( r)}{A^2}\dot A\,dt\equiv \\ \equiv-k(r) A+\alpha(r)+\int\frac{\gamma( r)}{A^2}dA\Rightarrow\\ \Rightarrow \left(\frac{\dot A}{A}\right)^2=-\frac{k(r)}{A^2}+\frac{\alpha(r)}{A^3}-\frac{\gamma(r)}{A^4}.
\end{multline} 
Let us define the following expressions:
\begin{align}
-k(r)&\equiv H_0^2(r)\,\Omega_k(r)\,A_0^2(r)\\
\alpha(r)&\equiv H_0^2(r)\,\Omega_m(r)\,A_0^3(r)\\
-\gamma(r)&\equiv H_0^2(r)\,\Omega_\gamma(r)\,A_0^4(r)
\end{align}
with the constraint $\Omega_k( r)+\Omega_\gamma( r)+\Omega_m( r)=1$, Eq.~\eqref{eq:fieldEquation} can be written in this form:
\begin{equation}
\frac{\dot A}{A}=H_0(r)\left[ \Omega_k(r)\left( \frac{A_0}{A} \right)^2+\Omega_m(r)\left( \frac{A_0}{A} \right)^3+\Omega_\gamma(r)\left( \frac{A_0}{A} \right)^4 \right]^{1/2}
\end{equation}
which can be directly integrated as follows:
\begin{equation}
\label{eq:cornerstone}
-H_0(r)\,t=\int_\frac{A(t,r)}{A_0}^1\frac{dx}{\sqrt{\Omega_k(r)+\Omega_m(r)x^{-1}+\Omega_\gamma (r)x^{-2}}}.
\end{equation}
Equation \eqref{eq:cornerstone} is the cornerstone of all the discussion: in fact it is possible to find the scale factor just by solving the integral and inverting the solution. However, the last step is not simple; in fact, if $\Omega_k( r)\ne 0$ there is no analytical way to invert the equality. Having this purpose in mind, let's impose $\Omega_k( r)=0$: this is not just a theoretical assumption due to the fact that the spatial curvature is constrained to be almost null by CMB observations. In such a way, we obtain that:
\begin{multline}
\int\frac{dx}{\sqrt{\Omega_mx^{-1}+\Omega_\gamma x^{-2}}}=\\=\frac{2}{\Omega_m}\int x\frac{d}{dx}\left( \sqrt{\Omega_\gamma+\Omega_mx} \right)\,dx=\\=\frac{2\sqrt{\Omega_\gamma+\Omega x}}{3\,\Omega_m^2}\left( \Omega_m\,x-2\,\Omega_\gamma \right).
\end{multline}
This result can be put into Eq.~\eqref{eq:cornerstone} and the result is the following:
\begin{multline}
\left[ -\frac{3\,\Omega_m^2\,H_0(r)\,t}{2}-\Omega_m+2\,\Omega_\gamma \right]^2=\left( \Omega_\gamma+\Omega_m\frac{A}{A_0} \right)\left[ \Omega_m^2\left( \frac{A}{A_0} \right)^2+4\Omega_\gamma^2-4\,\Omega_\gamma\Omega_m\left( \frac{A}{A_0} \right) \right]\Rightarrow\\ \Rightarrow \Omega_m^3\left( \frac{A}{A_0} \right)^3-3\,\Omega_m^2\Omega_\gamma\left( \frac{A}{A_0} \right)^2+4\,\Omega_\gamma^3-\left[ 2\,\Omega_\gamma-\Omega_m-\frac{3\,\Omega_m\,H_0\,t}{2} \right]^2
\end{multline}
%
Now the scale factor is given by the solutions of a third order polynomial equation. In particular, we consider the solution with $\dot A>0$, i.e.
\begin{equation}
A(t,r)=A_0(r)\left[ \frac{\Omega_\gamma(r)}{\Omega_m(r)}+\frac{\Omega_m(r)\,\Omega_\gamma^2(r)}{M(t,r)}+\frac{M(t,r)}{\Omega_m^3(r)}	\right]
\end{equation}
where
\begin{align}
M(t,r)&=\Omega_m^2\left( \frac{N+2\,\Omega_\gamma^3+\sqrt{N^2+4\,N\,\Omega_\gamma^6}}{2}\right)^{1/3}\\
N(t,r)&=\left[\frac{3\,\Omega_mH_0( r)\,t}{2}+\Omega_m-2\,\Omega_\gamma \right]^2-4\,\Omega_\gamma^3.
\end{align}
Our solution contains three free functions: $A_0( r)$, corresponding to the actual shape of the scale factor, $H_0( r)$ that represents the actual value of the Hubble constant in each point, and $\Omega_\gamma ( r)$ that is the density of the electromagnetic field. The density of matter $\Omega_m( r)$ is fixed by the relation $\Omega_m( r)=1-\Omega_\gamma( r)$. With the new expansion parameter given by Eq.~(4.11) it is possible to explore new phenomena in the Universe from an astrophysical point of view. 
\section{Electromagnetic bubble model}
In order to study the presence of electromagnetic field bubbles in the Universe, we have to fix the free functions $A_0( r)$, $H_0( r)$ and $\Omega_\gamma( r)$. First of all, we require $A_0( r)=r$; this choice means that distances at the present epoch ($t=0$) are simply evaluated as in the euclidean scenario.
Moreover, let us choice the following expression for the Hubble function:
\begin{equation}
H_0( r)=\bar H+\Delta H\exp\left( -\frac{r}{r_v} \right)
\end{equation}
where $\bar H+\Delta H$ is the Hubble constant evaluated at $r=0$, $r_v$ is the typical length at which the inhomogeneities can be appreciate and $\bar H$ is the value of the Hubble constant outside this region. Differently from the typical assumptions of an LTB inhomogeneous model, the void bubble is not real but it is a global manifestation of the superposition of a lot of small inhomogeneous regions. This effect can be appreciable at distance smaller than $r_v$.
This ansatz can mimic the dark energy effects as shown in \cite{cosmai}. In such a way, neglecting the dark energy component in the equations is a justified choice.
Now let us consider the presence of a single bubble in which an electromagnetic field is present. Clearly this context breaks the isotropy; however, if we assume that the intensity of the electromagnetic field is small enough, it is possible to use the isotropic solution, looking at the isotropy-breaking term simply as a `little blemish' of the solution. Having this feature in mind, we model our qualitative description by using a gaussian profile for the electromagnetic bubble,
\begin{equation}
\Omega_\gamma=\Omega_{\gamma0}\,e^{-\frac{(r-r_0)^2}{2\Delta r^2}-\frac{(\theta-\pi/2)^2}{2\Delta \theta^2}-\frac{(\phi-\pi)^2}{2\Delta \phi^2}}
\end{equation}
where the bubble is centered at $r_0$, $\phi_0=\pi$ and $\theta=\pi/2$, while $\Delta r$, $\Delta \phi$ and $\Delta \theta$ respectively represent the radial and the angular widths. However, $\Delta \theta$ and $\Delta \phi$ depend on $\Delta r$ and $r_0$: in fact, having in mind a spherical shape with radius $\Delta r$, the following relation is valid: $\tan(\Delta\theta)=\tan(\Delta\phi)=\Delta r/r_0$. In such a way, when $r_0\gg\Delta r$ let us approximate $\tan x\sim x$, i.e.
\begin{equation}
\Omega_\gamma=\Omega_{\gamma0}\,e^{-\frac{r_0^2}{2\Delta r}\left[ \left( 1-\frac{r}{r_0} \right)^2+\left( \theta-\frac{\pi}{2} \right)^2+\left( \phi-\pi \right)^2 \right]}.
\end{equation}
This is a new result in cosmology. In a next paper \cite{prep} we study the cosmological connections with astrophysical sources and we will consider new possible scenarios in the Universe.
\\
\\
\\
\section{Conclusion}
The result shows that the inhomogeneous cosmological background modifies the Maxwell equations as we can see in the Sect.~2. What distinguishes our approach is mostly its simplicity.
The expansion of the Universe is determined by the scale factor $a(t)$, in which $H(t) = \dot{a} / a$ is the Hubble expansion rate.
The presence of the inhomogeneity with electromagnetic field gives rise to a new scale factor given by eq.~(4.11).
It is important to stress that inhomogeneities may have important effects on the propagation of photons in the Universe and this may be important making observations; in fact observers obtain information about the Universe by means of photons, see for example \cite{{fuma},{brouzakis},{meures}}.
\\
The dichotomy consists in the apparent homogeneity of the Universe, while it is also well fitted by an  inhomogeneous evolution of the same Universe. 
\\
We have also discussed the possibility of 
an electromagnetic field bubbles in the Universe. Taking into account the expansion for the Hubble function given by Eq.~(5.1) we have obtained the most general $\Omega_{\gamma}$. This may be studied further by means of future cosmological observations. Our interesting lesson from the consideration of this paper is that if one consider electrodynamics in curved space-time with a LTB inhomogeneous Universe it is possible to determine a very interesting expansion factor of the Universe. In any case , whatever direction the study of LTB cosmologies might take, results of this investigation should be relevant.
\\ 
The results presented in this article will be helpful to analyze experimental effects in a more actual inhomogeneous cosmological scenario, and we are presently attempting to generalize our results. Work still has to be done in order to explore these new cosmological effects,  but this is a story for another work.
\end{document}